\documentclass[msmath,amssymb,aps,prl, twocolumn, epsfig, showpacs, bibliography, lengthcheck]{revtex4-1}

\usepackage{graphicx}% Include figure files
\usepackage{subfigure}
\usepackage{dcolumn}% Align table columns on decimal point
\usepackage{bm}% bold math
\usepackage{hyperref}% add hypertext capabilities
\usepackage{epstopdf}
\usepackage{animate}

\begin{document}
\title{Spectra Analysis to Stretching of ADB Structure Metamaterial}% Force line breaks with \\
\author{Q. Sun, J.-Y. Tang, J.-N. Lin, M.-G. He, Z.-P. Wang}
\affiliation{Department of Physics, University of Science and Technology of China, Hefei, Anhui, China 230026}

\date{\today}
\begin{abstract}
Asymmetric-double-bars (ADB) structure is one of the most interesting plasmonic metamaterials that has been broadly investigated. Here we propose to manufacture ADB on top of elastic material, to get direct control to the dimension of ADB elements. To analyze the spectra numerically, simulation by commercial software (COMSOL) are carried out. We successfully modify the characteristic spectra and enhance Q-factor of the peak near infrared by introducing angular and amplitude parameters of the stretching of substrate in the simulation. At the mean time, we significantly restrain red shift in the absorption spectra by applying flipped-configuration and substrate etching configuration to ADB structure. Intriguing quadratic functions between stretching ratio and the absorption peak wavelength are obtained when stretching in x and y direction. For other directions, EIT lineshape appears in transmission spectra. These results might contribute to future application of plasmonic metamaterial in laser controlling and sensors.
 \end{abstract}
%\pacs{32.30.Dx, 51.20.+d, 06.30.Gv}

\maketitle

%这部分是引言，需要照顾到下面的几个方面
In nature, most plasmonic materials don't have strong interaction with the magnetic component of light. With the development of nanotechnology, plasmonic metamaterial, usually formed by metal array in nanoscale, came into sight and rapidly developed for its artificial magnetism at optical frequencies. With further researches, plasmonic metamaterial gives out many possible intriguing properties, such as negative refraction index material \cite{negative1,negative2,negative3,negative4}, extraordinary transmission \cite{trans1,trans2}, and a lot more possibilities. Much effort has been made to explain this phenomenon. The widely accepted explanation to these features is the surface plasmons (SPs) theory \cite{wikisps}. Studies have shown that metals such as Au and Pt \cite{EIT7} are unique in that they can enhance the magnetic field near the surface, resulting in some characteristic spectrum.

Many possible configurations of plasmonic metamaterial have been developed, for example split rings \cite{split1,split2,split3,split4}, has been found related with intriguing properties. Another important structure is asymmetric-double-bars(ADB)\cite{adb1}, which is mainly investigated in this paper. One element of ADB matrix consists of two bars, slightly different in length. It is one of the simplest structures in metamaterials, resulting in relatively easy manufacture and simulation. ADB is also promising for realizing sharp Fano resonance \cite{fano1,fano2,fano3}. A quadrupole-like dark mode with small radiative loss is excited by a free space electromagnetic wave because of asymmetry of the ADB. It consists of two dipoles with opposite phases, and Fano interference occurs between the quadrupole mode and the dipole mode. A net dipole moment is small, which leads to weak radiative loss and high Q-factors. Therefore, ADB structure is applicable in many occasions, for example, possibility of electromagnetically induced transparency (EIT) \cite{EIT1,EIT2,EIT3,EIT4,EIT5,EIT6,EIT7}, and modification to the fluorescence spectrum of coupled Quantum Dots \cite{OMwithQD}.

\begin{figure}
  \centering
  \includegraphics[width=3in]{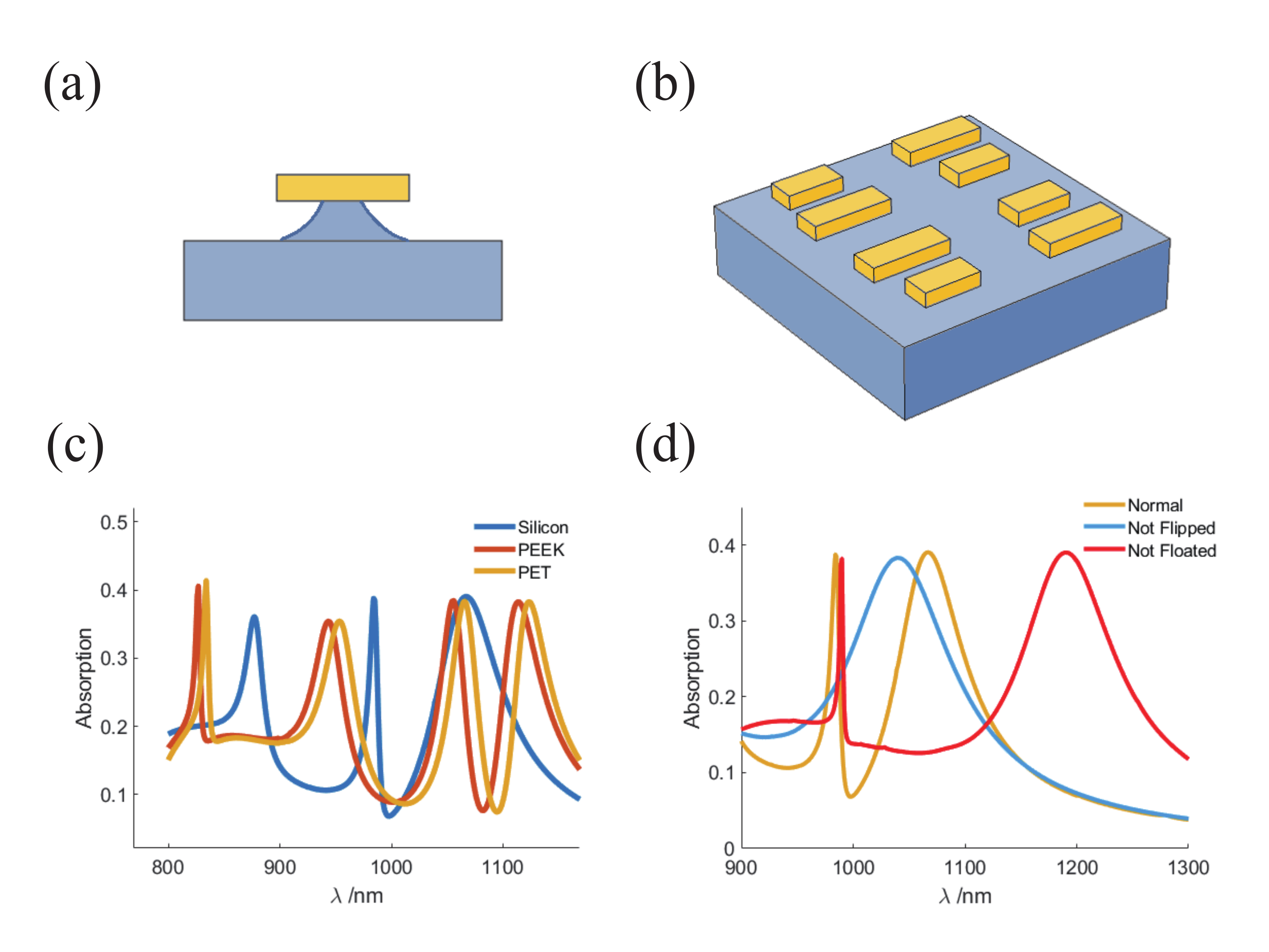}
  \caption{\label{fig:structure}(Color online) (a) Substrate etching diagram, and the connection is omitted in simulation; (b) Demonstration of simulation model -- ADB structure on surface of silicon dioxide or plastic; (c) Absorption spectra of same ADB structure on different substrate, silicon dioxide, PET and PEEK； (d) Silicon dioxide absorption spectra of alternately flipped+substrate etched, non flipped+etched, and flipped+unetched, respectively.}
\end{figure}

%flipped
In our model, we use ADB formed by gold for high electrical conductance. It has been found out that alternately flipping the ADB structure, that is to make the adjacent elements to be head-to-head rather than head-to-tail, can enhance the quality factor of the absorption spectra, according to previous work by Yuto Moritake et al. \cite{anti}. Since the dipole-dipole interaction in the non-flipped structure can be undesirably affected from outside of the periodic structure, alternately flipping can prevent this and enhance the should-be interaction and increase the Q-factor. Another problem in nanostructure fabrication is to reduce the influence to the electric and magnetic field by the substrate. The reduction of substrate influence can be achieved by using selective and isotropic etching of the substrate as demonstrated in \cite{foat}. Substrate etching was applied to a silicon substrate under the gold nanostructures shown as FIG.\ref{fig:structure}(a). And it was experimentally proved that it can prevent the red shift and improve the refractive index sensitivity based on far field interference. However, the decrease in the Q-factors of the Fano resonance is observed after applying substrate etching, which is due to the electric field distortion by a closely placed substrate. In our experiment, we combine the alternately flipped configuration of ADB and substrate etching together in order to achieved the finest spectral properties of Fano resonance.

%这部分介绍大小
The schematics of configuration is shown as FIG.\ref{fig:structure}(b). The unit cell of the ADB is composed of two gold bars of which the size is \emph{l1}=280 nm, \emph{l2}=200 nm, and width \emph{w}=100 nm, gap between two bars \emph{w}=100 nm, height \emph{h}=50 nm. The periodic lengths in x,y directions are both 500 nm, and four alternately-flipped ADB units are placed symmetrically in four quadrants of square. Substrate etching is also applied in our design, which is estimated to be 25 nm during simulation. In experiment, we suggest coating another 5 nm Ti between gold bar and the silicon dioxide to improve the adhesive strength, which is not shown in this figure.

%使用塑料的好处
Previous fabrication of ADB metamaterial was always conducted on silicon dioxide, because of its high transparency to all colors of light, endurance to laser and stable chemistry property. Yet silicon dioxide retains the possibility of being elastic to change the period and the distance, which means that one piece of sample corresponds to one spectra. In order to improve the flexibility of the experiment, the substrate will have to meet the requirement mentioned above, and have the controllable flexibility. Piezoelectric material and thermal expansion material share the flexibility but the degree of relative deformation of the former is too small and the direction of expansion of the latter can hardly be controlled. Thus, we try to achieve this goal by mechanically manipulating suitable elastic plastic. According to the description above, we think that Polyethylene Terephthalate (PET) and Polyether Ether Ketone (PEEK) are both worth expecting candidates. However, it might require more stringent conditions to EBL process, which will not be discussed any further in this paper.

\begin{figure}
  \centering
  \includegraphics[width=3in]{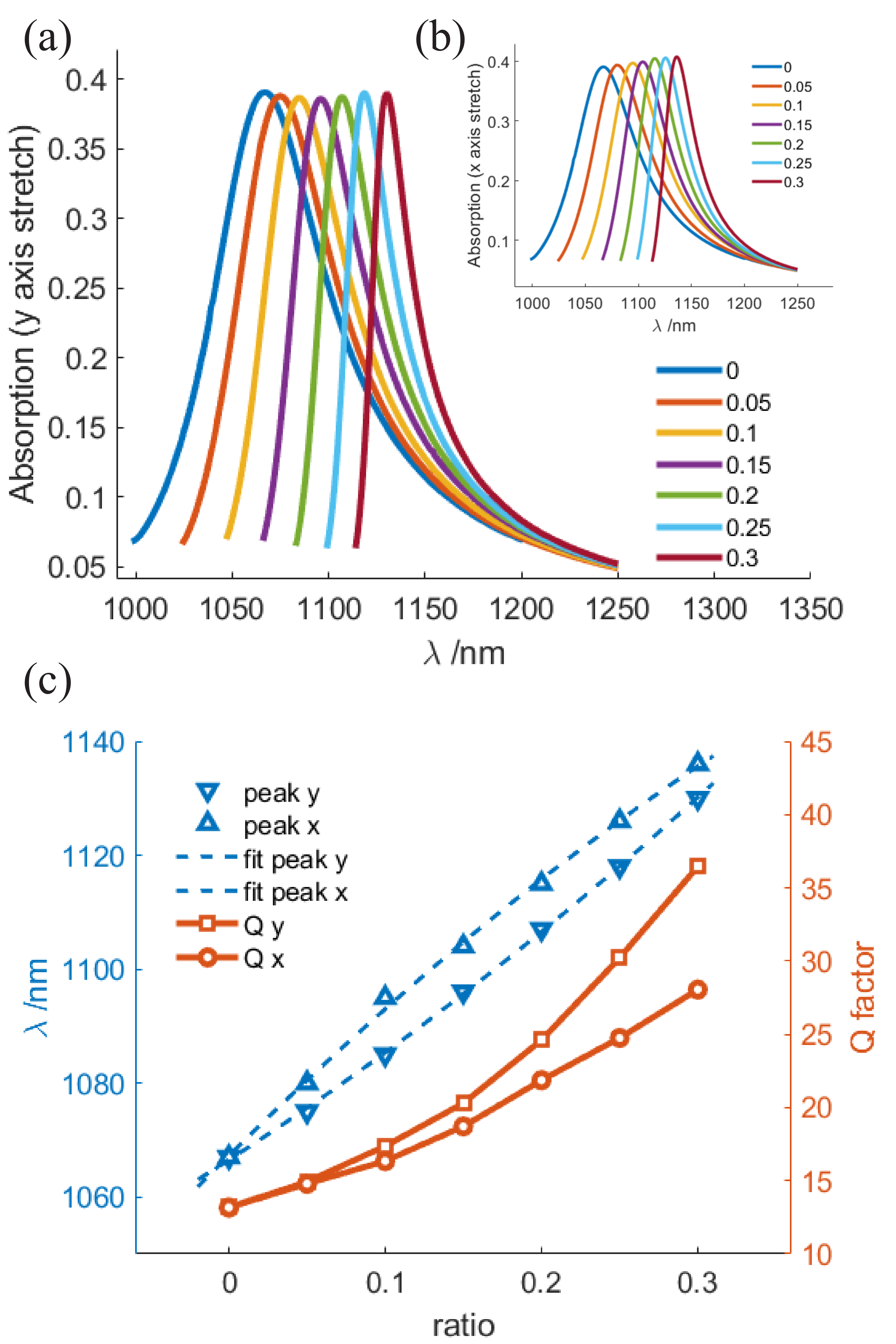}
  \caption{\label{fig:xy}(Color online) (a) Simulation result of absorption spectra with different \emph{Ratio} in y and (b) x axis on silicon dioxide; (c) wavelength and the Q-factor of \emph{R peak} relative to \emph{Ratio}, and the fitting result of peak to \emph{Ratio} using quadratic function. The fitting residual norm of peak y is 1.0465, for peak x is 2.3805.}
\end{figure}
%制备流程

%材料区别
Here we present our simulation result based on the finite element method (FEM) by commercial software COMSOL. As is shown in FIG.\ref{fig:structure} (c), we perform spectra analysis of the designed ADB structure on different substrates, including silicon dioxide, PET and PEEK. There are three absorption peaks for silicon dioxide. We observe the spatial magnetic field around ADB and find the interaction between dark mode and bright mode similar with each other. Four peaks for PET and PEEK are observed, and it looks like the two peaks on the right derive from the right one of silicon dioxide spectra. Considering the difference on reflective index molecular structure, it's not surprising to see the difference. And it is possible to make use of these two peaks for higher Q-factor. To clarify, we name the right peak on the silicon dioxide absorption spectra to be \emph{R peak}. What's important is, we observe the similar reaction of two-peaks (from PET and PEEK) compared with \emph{R peak} in the stretching process, including the separation of peaks, which means the one-peak becomes two-peak, and the two-peak becomes four-peak. We will come to this part soon. Hereby, we will take more look into to the \emph{R peak} on silicon dioxide spectra.

\begin{figure}
  \centering
  \includegraphics[width=3in]{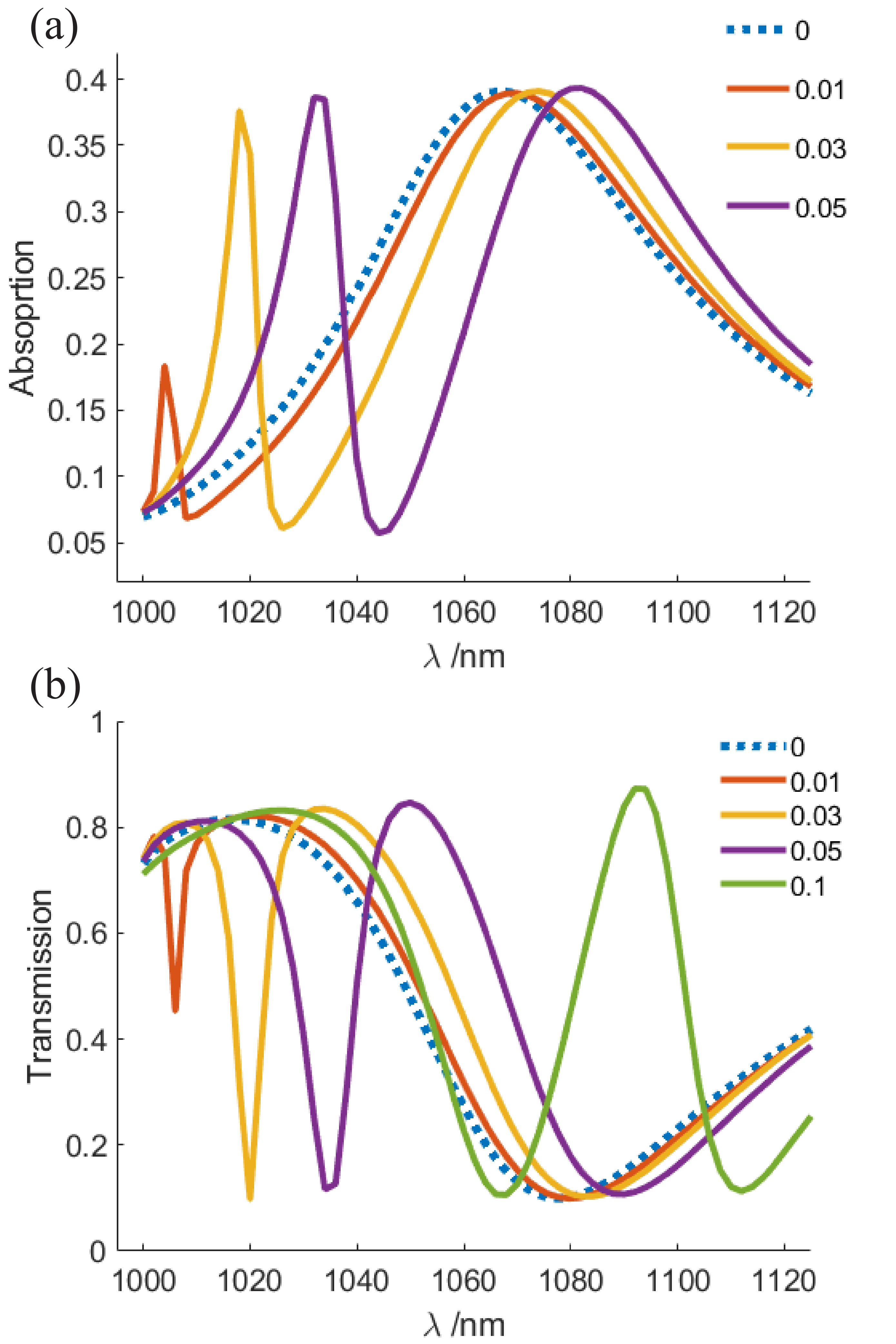}
  \caption{\label{fig:eitline}(Color online) (a) when $\theta$ is set to $\pi/4$, and scan \emph{Ratio}, then a small peak in the absorption spectra grows; (b) the corresponding transmission spectra, and an EIT lineshape appears because of the emergence of a metastable state.}
\end{figure}

%翻转什么的,颜色改一下
As shown in FIG.\ref{fig:structure} (d), we also run simulation to prove the reasonability of alternately flipped configuration and substrate etching. Yellow line refers to alternately flipped configuration with substrate etching, and red line and blue line refer to alternately flipped configuration without substrate etching and non-flipped configuration with substrate etching, respectively. It is observed from the spectra that flipping unit cells would create a sharper peak and substrate etching would eliminate red shift. Alternately flipped configuration with substrate etching, which combined both advantages, would have a sharper peak than non-flipped configuration with substrate etching and much weaker red shift compare with alternately flipped configuration without substrate etching.

%x y
In order to characterize the stretch, we set two parameters in the model. $\theta$ means angle between the stretching direction and x axis, and \emph{Ratio} means the rate of increase in the specific direction. First of all we can see the spectra modification when $\theta=\pi/2$ or 0, that is y and x direction, as is shown in FIG.\ref{fig:xy} (a) and (b), which means in y and x axis, respectively. Different lines represents different \emph{Ratio}. And we can see the \emph{R peak} tends to be continuously red shifting, and FWHM increasing, as \emph{Ratio} grows. We can consider the enlargement of space between the two bars is like that of an optical cavity. By pulling the two bars away from each other, the resonance wavelength is to red shift. Here the effect of head-to-head structure is also weakened, thus the red shift isn't suppressed.

%fit结果
To describe the red shift and change in FWHM, we find peaks shown in FIG.\ref{fig:xy} (a) and (b) and plot them with \emph{Ratio}, which is shown in FIG.\ref{fig:xy} (c). Y-axis on the left refers to \emph{R peak} wavelength, on the right is the Q-factor, and $Q=\frac{\lambda_{peak}}{\Delta\lambda}$, where $\Delta\lambda$ is FWHM. Monotone increasing is observed. Then we use the quadratic function to fit the curve of the peak to ratio, as is also shown in figure FIG.\ref{fig:xy}. In order to explain that, we can regards ADB structure as an optical cavity, and the eigen-wavelength should be proportional to cavity length as a result. And we can see that when \emph{Ratio} is from 0 to 0.2, linear distribution is obvious; yet other factors will introduce higher order terms, as the quadratic function we use to fit indicates second-order term. These nonlinear terms come from various causes, for example, the impact caused by flipped structure should decrease in a quadratic way as \emph{Ratio} goes up, and if the double bars get too close or too far-away, it will also introduce nonlinear term. The second-order approximation can work fine when \emph{Ratio} is smaller than 0.3, which is most likely to happen in future applications.

\begin{figure}
  \centering
  \includegraphics[width=3in]{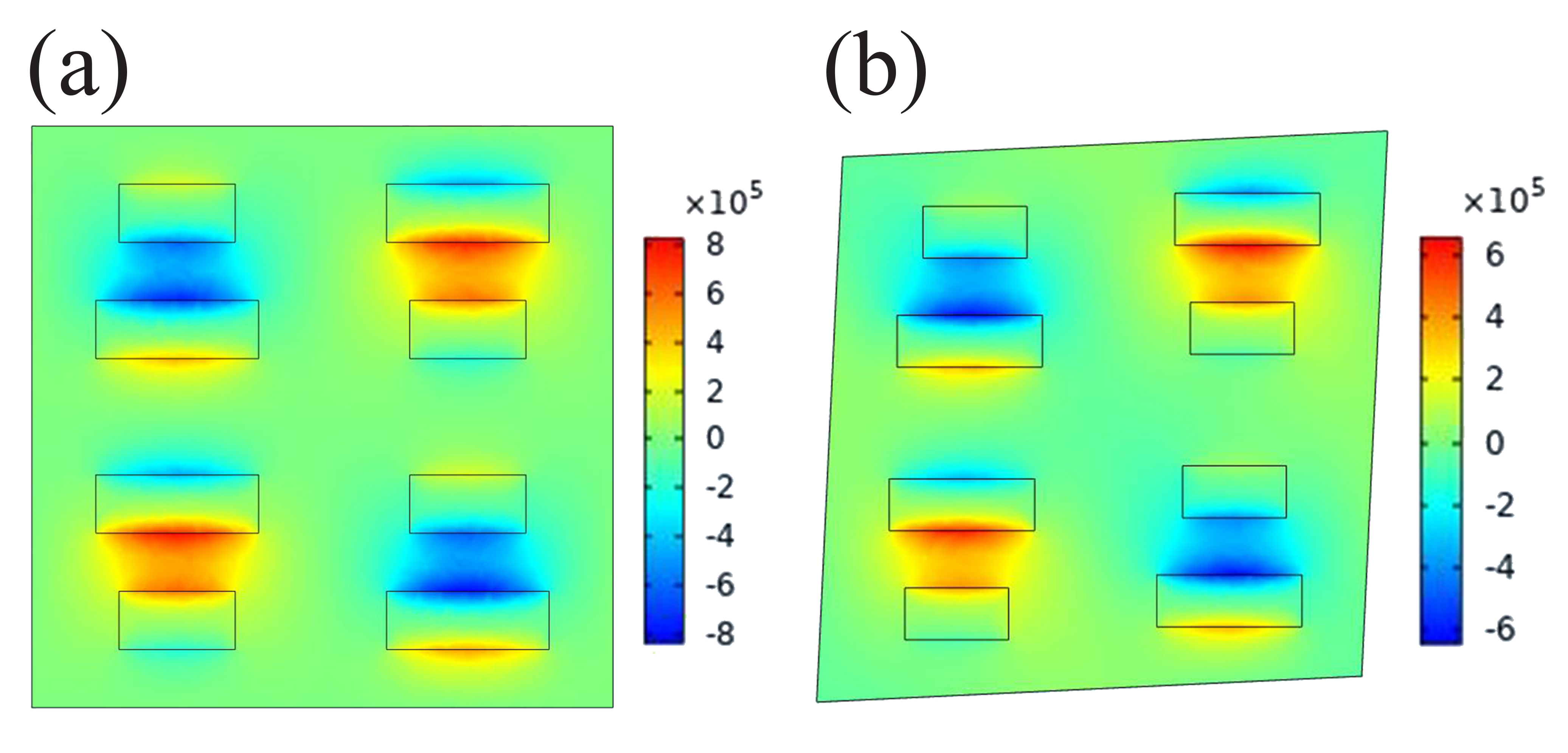}
  \caption{\label{fig:Hz}(Color online) (a) The spatial distribution of Hz when \emph{Ratio}=0, the corresponding wavelength is 1065 nm, Hz is in unit of A/m, which is not shown in figure, and strong resonance appears; (b) The spatial distribution of Hz when \emph{Ratio}=0.1, $\theta=\pi/4$, the disruption of structure symmetry and Hz distribution symmetry is observed.}
\end{figure}

We also study on different angles, for example $\theta=\pi/4$, and interesting modification is discovered. If we set the parameter $\theta=\pi/4$, and scan \emph{Ratio}, an EIT lineshape appears, as in FIG.\ref{fig:eitline} (b). When \emph{Ratio} is small, a small peak in the absorption spectra, near \emph{R peak}, begins to grow with \emph{Ratio}. As \emph{Ratio} becomes big enough ($\approx 0.1$), the little peak becomes as big as \emph{R peak}. And if we see FIG.\ref{fig:eitline} (a), the transmission spectra, the characteristic lineshape of EIT is shown. We may consider the small peak help generate a metastable state near the fano resonance peak, and the small peak comes from the disruption of symmetry between two bars. ADB structure in our design have an axis of symmetry. And stretching in an angle like $\pi/4$ breaks this sysmmetry, and introduce a bias of the bright mode in spatial distribution, which has influence on the coupling between the bright mode and the dark mode. According to FIG.\ref{fig:Hz}, the spatial distribution of Hz have changed. Moreover, the peaks continues to red shift with the growing of \emph{Ratio}, which is estimated before. If we set \emph{Ratio} constant, and scan $\theta$, it will introduce more nonlinear effects, which makes the analysis less applicable. However, when \emph{Ratio} is 0.1 and nonlinear effects are unnoticeable in this particular region, it seems that the transmission has give out some hints for future investigation. And in other region, nonlinear effects are strong, which makes the scanning of $\theta$ less useful. But when $\theta$ is close to $\pi/2$ or 0, the difference in spectra is ignorable, which means it's not to sensitive to direction error in future application.

\begin{figure}
  \centering
  \includegraphics[width=3in]{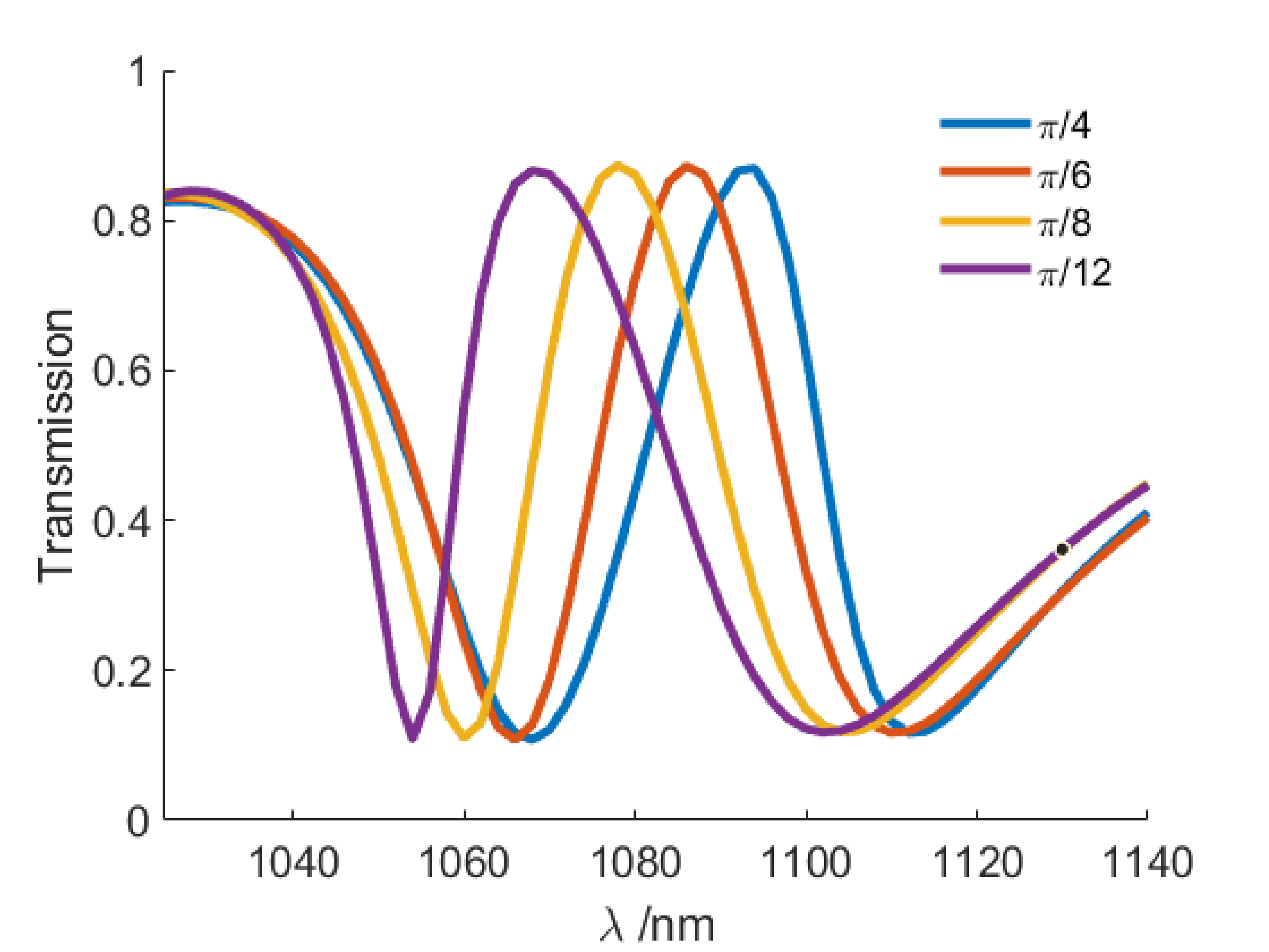}
  \caption{\label{fig:theta}(Color online) The transmission spectra when \emph{Ratio}=0.1, and $\theta$ is scanned. This peace is trivial since the strong nonlinear effects would appear in other region.}
\end{figure}

In conclusion, plasmonic metamaterial is becoming more and more important in scientific investigation for its potential in improving quality factor of sensors and quantum dots and so on. Adjusting the plasmonic metamaterial in ADB structure with one piece is tempting, which might bring great effect to this industry. We run simulation on silicon dioxide and set two important parameters, \emph{Ratio} and $\theta$, to represent the stretching on the substrate. When $\theta$ equals to $\pi/2$ or 0, the \emph{R peak} wavelength is the function of \emph{Ratio}, which is well fitted by quadratic function. An increasing Q-factor is also observed. When we set $\theta$ to be $\pi/4$, and scan \emph{Ratio}, EIT lineshape in transmission spectra appears for the breaking of structure symmetry. This phenomenon is also observed when $\theta$ is at other values. And when we set \emph{Ratio} constant and scan $\theta$ the other way, nonlinear effects will take over and when $\theta$ is small, the effects are negligible. All phenomenon mentioned above are seen in PET and PEEK spectra as well. By doing so, we see the chance of manipulating plasmonic metamaterial by using elastic substrate. They could have various applications, for example, tuning the emission wavelength of fluorescence by quantum dots like PbS and graphene, and the possibility of willingly adjustment of near-monochrome beam.

We would like to thank researcher Kun Zhang and Dianfa Zhou from USTC Center for Micro- and Nanoscale Research and Fabrication, for their experimental help in our fabrication of plasmonic metamaterial; we also thank Prof. Zhongping Wang for his instructions throughout our experiment. We acknowledge Tanwei Li and Jian Zuo from USTC Physical and chemical science laboratory center for providing consultation to our experiment.

\end{document}